\documentclass[a4paper]{jpconf}
\usepackage{graphicx}
\usepackage{textcomp}
\begin{document}
\title{TPC development by the LCTPC collaboration for the ILD detector at ILC}

\author{Jochen Kaminski for the LCTPC collaboration}

\address{Rheinische Friedrich-Wilhelms-Universit\"at Bonn, Physikalisches Intitut, Nussallee 12, 53115 Bonn, Germany}

\ead{kaminski@physik.uni-bonn.de}

\begin{abstract}
  The International Large Detector (ILD) at the International Linear Collider (ILC) requires significantly improved subdetector systems to comply with the envisioned performance. Its central tracking detector is a Time Projection Chamber (TPC) which is equipped with Micropattern Gaseous Detector (MPGD) readout modules. The LCTPC collaboration was founded in 2007 to design such a MPGD-based TPC and to build and test demonstrator modules. For this an experimental setup at DESY, Hamburg, was built, where four different readout concepts have been studied in many successful test beam campaigns since 2010. It could be demonstrated that the required transverse and longitudinal space point resolutions can be reached. Also, a new gating concept was developed that can effectively eliminate the ion backflow, while marginally degrading the performance of the detector.  

%  An extensive research and development program for a ILC TPC has been carried out within the framework of the LCTPC collaboration. A Large Prototype TPC in a 1 T magnetic field, which allows to accommodate up to seven identical Micropattern Gas Detector (MPGD) readout modules of the near-final proposed design for the ILD detector at ILC, has been built as a demonstrator at the 5 GeV electron beam at DESY. Three MPGD concepts are being developed for the TPC: Gas Electron Multiplier, Micromegas and GridPix. Successful test beam campaigns with different technologies have been carried out between 2014 and 2019. Fundamental parameters such as transverse and longitudinal spatial resolution and drift velocity have been measured. In parallel, a new gating device based on large-aperture GEMs have been produced and studied in the laboratory. In this talk, we will review the track reconstruction performance results and summarize the next steps towards the TPC construction for the ILD detector.
\end{abstract}

\section{Introduction of the Project}
The International Linear Collider (ILC) is one of the possible future projects proposed by the HEP community. It is a linear electron positron collider with center of mass energies increasing in several stages from 250 GeV possibly up to 1 TeV with the possibility of short calibration runs at 91 GeV. One of the main goals is to study the Higgs-boson and determine its properties with high statistics and high precision. To record the events two experiments are foreseen to be exchanged at the interaction point by a push-pull setup. One of the two experiments is the International Large Detector (ILD), which is described in Ref. \cite{ILD}. The design idea of the experiment is the particle flow concept, which was developed to optimize the physics output of HEP experiments by using high granularity trackers and calorimeters to reconstruct the events with the best possible momentum and energy resolution. A Time Projection Chamber (TPC) was chosen as a main tracking device, because several of its features are important to the overall detector concept. Most notably the large number of true 3D space points along each track facilitate track finding and reconstruction with high efficiency and therefore are important ingredients for realizing the particle flow concept.

 Performance requirements of the TPC have been documented in \cite{ILD} and surpass the requirements of other large volume TPCs by a large factor. For example a momentum resolution of $\delta\left(1/p_t\right)\approx 10^{-4} \left(\mathrm{GeV}/\mathrm{c}\right)^{-1}$ is required for the TPC alone. It was shown that these requirements could not be reached in a high magnetic field of $B = 3.5~\mathrm{T}$ with a conventional wire readout. Thus, the LCTPC collaboration was formed in 2007 to develop a TPC with Micropattern Gaseous Detector (MPGD) readout modules. Currently, four different technologies are investigated and their performances are compared.
 
 \section{Test Setup at DESY}
 To reduce the redundancy in multiple setups, it was decided to construct a common test facility at DESY, where an electron test beam with momentum between 1 and 6 GeV/c is available. Some benefits of a common infrastructure are a common environment which facilitates the comparison of the different technologies. In addition it allows for a larger setup with seven readout modules and therefore to study integration issues such as field distortions between the readout modules or development of realistic module mounting procedures.

\begin{figure}[h]
\begin{minipage}{16pc}
\includegraphics[width=16pc]{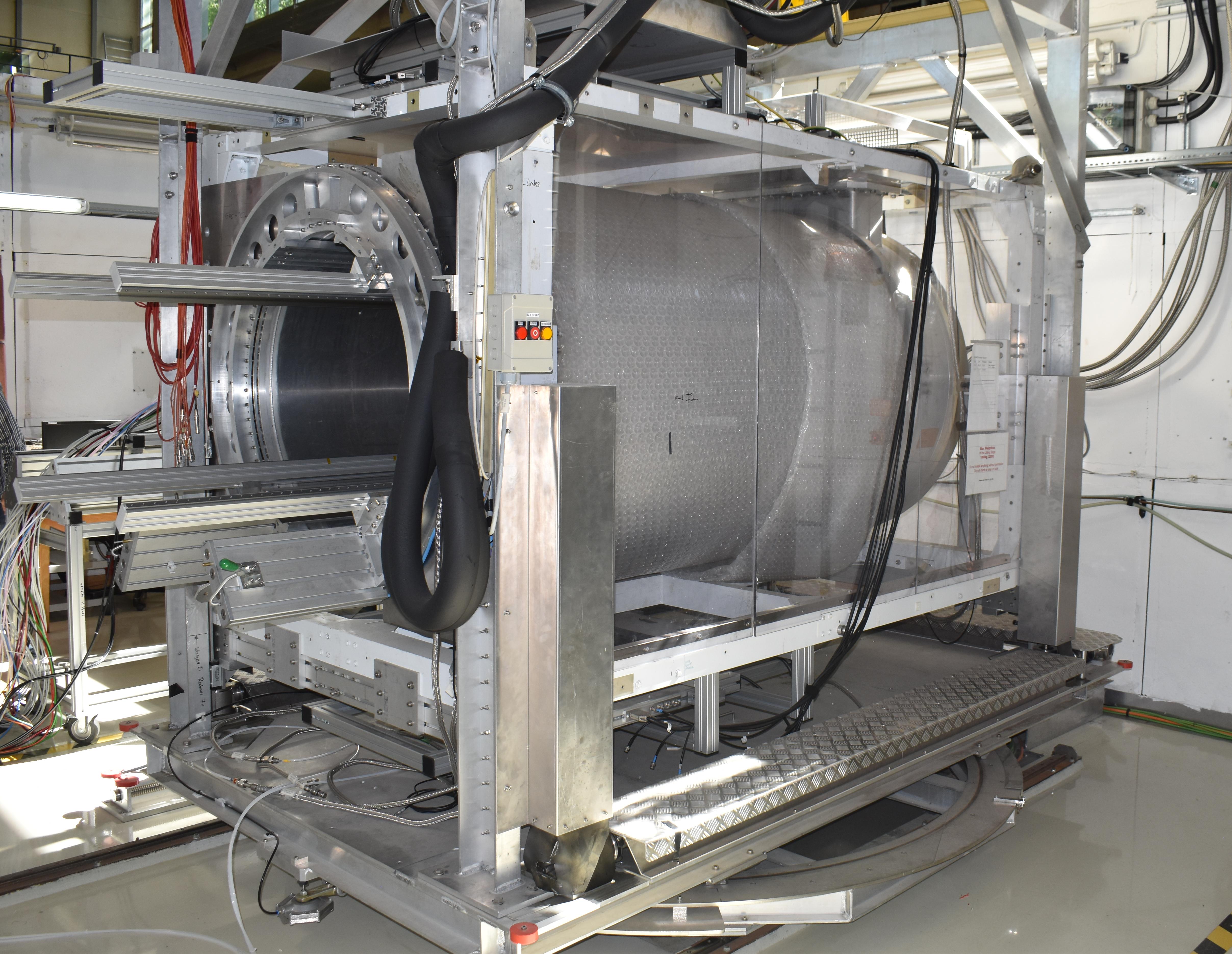}
\caption{\label{l_PCMAG}Test setup at DESY with the magnet PCMAG on a moving stage.}
\end{minipage}\hspace{2pc}%
\begin{minipage}{20pc}
  \vspace{-1pc}
  \includegraphics[width=20pc]{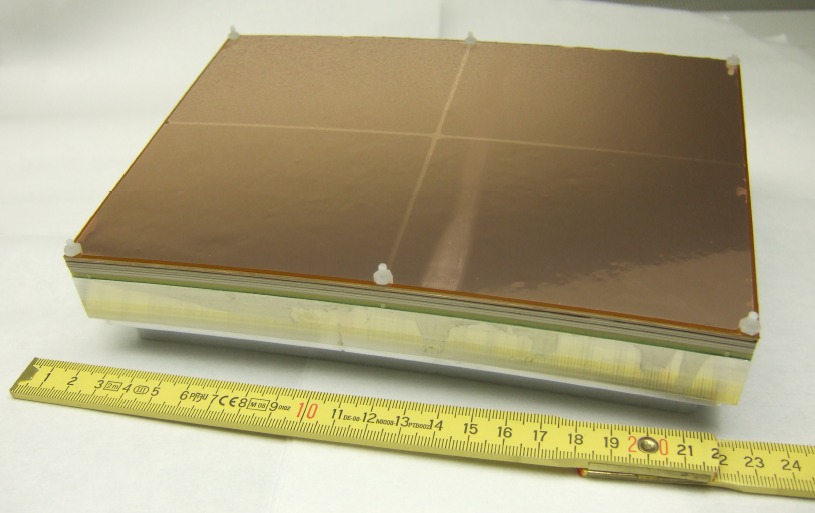}
\caption{\label{l_tGEM} Picture of a triple GEM module. \cite{PhDFelix}}
\end{minipage} 
\end{figure}

 The setup therefore comprises of the following components: A large superconducting magnet (PCMAG) with a magnetic field of up to $B = 1.2$ T is mounted on a movable stage (see Fig. \ref{l_PCMAG}), which allows for displacements in the two directions perpendicular to the beam and around a vertical rotation axis. A cylindical TPC prototype (Large Prototype, LP, \cite{LP}) with a maximum drift length of 61 cm and a diameter of 72 cm can be mounted inside the magnet and rotated around its central axis. The endcap of the TPC allows for up to seven readout modules roughly of the size $22 \times 17~\mathrm{cm}^2$, which can be replaced by dummy modules if needed. Additionally, an infrastructure for operating gaseous detectors, such as gas cabinets with instrumentation and piping, high voltage power supplies, beam and cosmic ray triggers based on plastic scintillators, 10,000 channels of readout electronics \cite{ALTRO} based on the ALTRO ASIC and a 2-phase CO$_2$ cooling plant for cooling tests of the electronics are available.

 The setup is constantly maintained and enhanced, for example the latest addition is the beam telescope LYCORIS \cite{LYCORIS}, which is mounted inside the PCMAG around the LP allowing for precise particle tracking within the magnet and thus avoiding the multiple scattering effects in the magnet walls ($2\times 0.2$ X$_0$). It can reach a point resolution of about 7 \textmu m per plane in the direction perpendicular to the beam. With this performance the momentum of particles can be determined event-by-event and the required momentum resolution of the TPC can be verified.

 \section{Readout Technologies}
Several different module design ideas are under evaluation experimentally and by simulation. To optimally implement these designs, different readout techniques and gas amplification stages have been chosen. In this section the different design ideas are discussed along with other design choices.  
 
\subsection{Triple GEMs}
The design idea of this module is to minimize the inactive area. To reach this standard gas electron multipliers (GEMs) with an insulation layer of 50 \textmu m of Kapton and thin copper layers as electrodes were chosen. These GEMs are glued onto thin ceramic frames, which are only 1~mm wide (see Fig. \ref{l_tGEM}). To increase the operational stability a triple GEM stack is used and the readout pads are rectangular and staggered with a pitch of $1.26\times 5.85$ mm$^2$. A first set of modules had been tested in the setup in 2014 and results have underlined that the requirements of the ILD detector can be met \cite{LCTPCGEM}. In a second iteration a more automatized production procedure including gluing has been used to achieve better and more reproducible results. A test beam campaign in 2016 could confirm the good performance of the module design and a publication including dE/dx measurements and double track resolution is in preparation \cite{LCTPCGEM2}. 
 
\subsection{Double GEMs}
\begin{figure}[h]
\begin{minipage}{16pc}
\includegraphics[width=16pc]{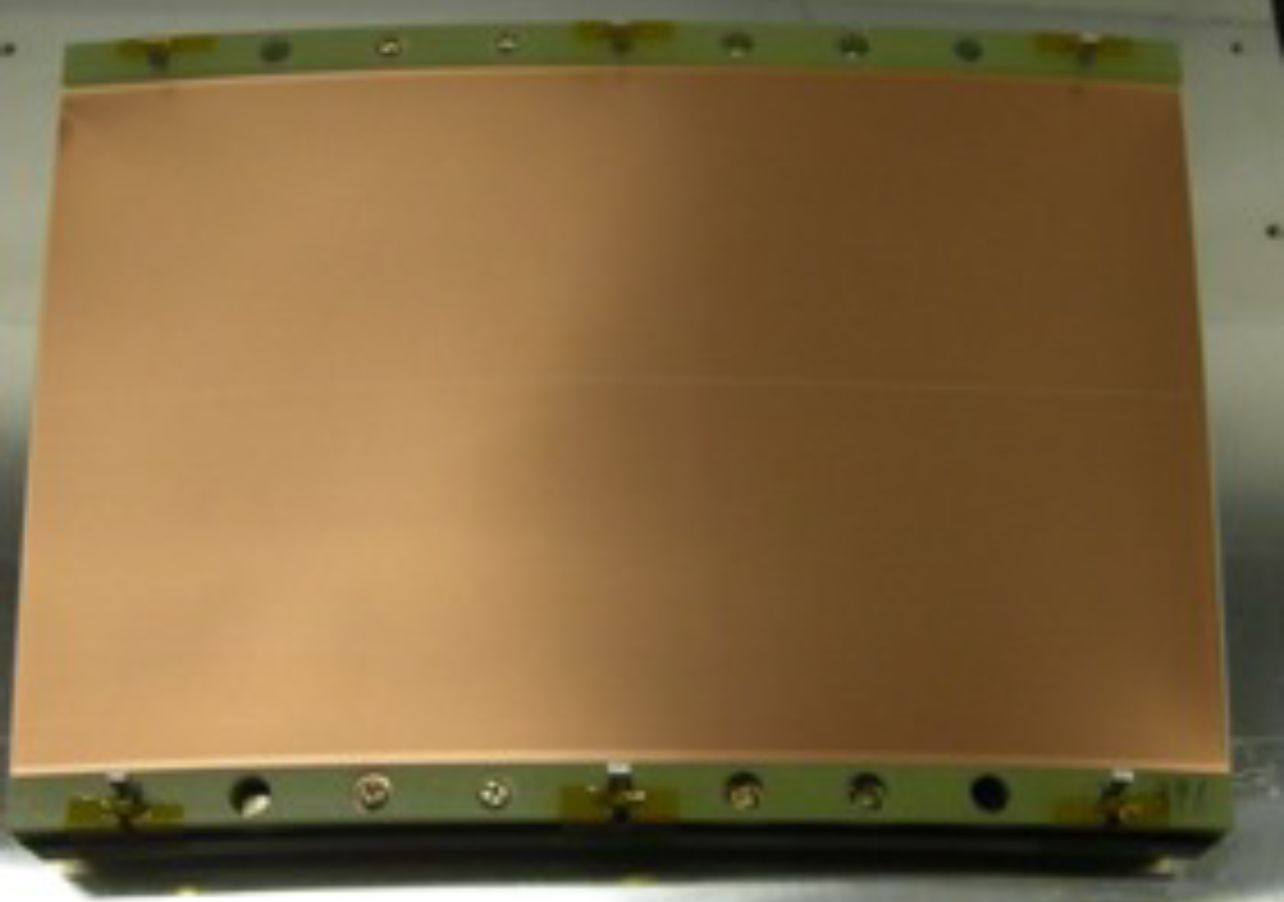}
\caption{\label{l_dGEM}Picture of a double GEM module.}
\end{minipage}\hspace{2pc}
\begin{minipage}{19pc}
  \vspace{-1pc}
  \includegraphics[width=19pc]{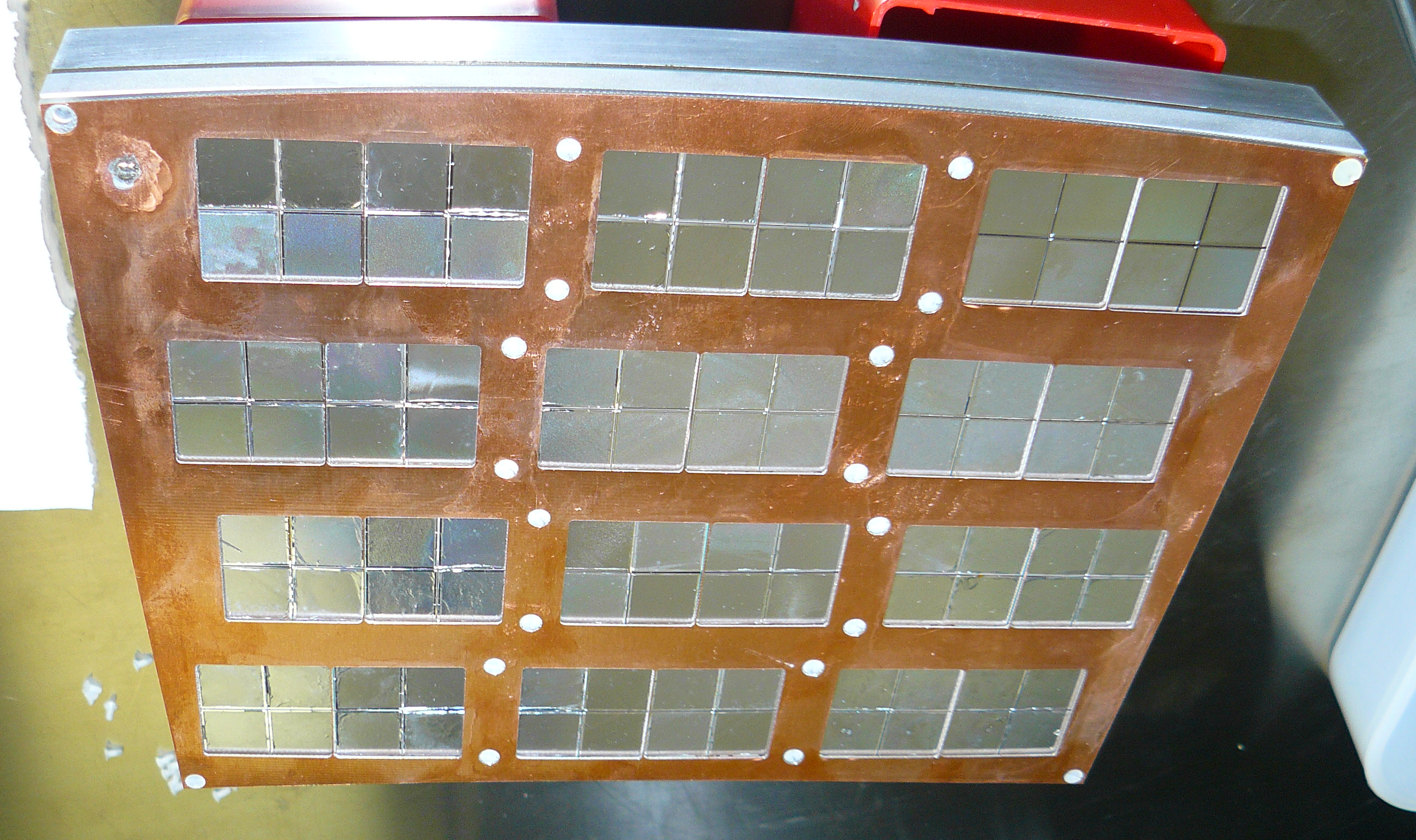}
\caption{\label{l_GridPix}Picture of a GridPix module.}
\end{minipage} 
\end{figure}

An alternative design idea is to minimize the dead area pointing towards the interaction point. For this no frames at the sides pointing in the $r$-direction are used, but instead the GEMs are stretched only with two thicker bars at the curved sides of the module (see Fig. \ref{l_dGEM}). To stabilize the GEM at the open sides, they are made with a 100 \textmu m thick insulator of liquid crystal polymer (LCP) \cite{AsianGEM}. As the higher thickness favors higher gas gains, only two GEMs are needed. The pads with a  pitch of $1.2\times 5.4$ mm$^2$ are also rectangular and staggered. The results show a very similar performance as the triple GEM modules. 

\subsection{Resistive Micromegas}
To cover a larger area with fewer and thus larger pads without degrading the performance a new readout concept was developed for the ILD-TPC: a resistive layer on the pads is spreading the narrow charge distribution of a Micromegas gas amplification stage over several pads thus enabling a more precise charge interpolation than otherwise possible \cite{rMM}. This concept allows for rectangular and non-staggered pads with a size of $3\times 7$ mm$^2$ to give similar results as the two GEM-type modules. A detailed study with different resistive materials and various resistive values have been performed first \cite{MM-TB}, then a small series production of seven modules was done to test a full coverage of the LP and demonstrate the technological readiness for mass production. In a final test, an inverted HV scheme was tested, where the amplification mesh is on ground potential and the resistive layer on a positive high voltage. This configuration showed a reduction of field distortions between the modules by one order of magnitude and is therefore favored for further studies.

\subsection{GridPixes}
To reach the best possible performance, GridPixes \cite{GridPixTP} have been chosen as a fourth option. A GridPix consists of a highly pixelized readout ASIC, in our application a Timepix ASIC with a pixel pitch of $55\times 55$ \textmu m$^2$ is used. The pixel area of the ASIC ist first covered with a resistive layer to protect the ASIC from discharges and a Micromegas is built on top with photolithographic postprocessing techniques. Since the charge avalanche created by a single electron is collected and registered by a single pixel, one can therefore conclude that every hit corresponds to a primary electron. This concept offers the best possible resolution only limited by the diffusion in the drift region.
As the active area of a single chip is about 2 cm$^2$, a large number of GridPixes is needed to cover the endcaps of a TPC. In a first demonstrator experiment 160 GridPixes were assembled on three modules and operated in the DESY setup for two weeks \cite{160GridPixTP}. The next development was the transfer from the Timepix to its successor the Timepix3 ASIC. Successful test beams with a single Timepix3 GridPix \cite{singleGridPixTP3} and a four GridPix device \cite{quadGridPixTP3} have demonstrated the potential of these detectors as far as spatial resolution and dE/dx resolution are concerned. A 32 GridPix detector is at the time of writing being tested inside PCMAG at the DESY test beam.

\section{Suppression of Ion Backflow}
Simulations have shown that the expected beam conditions and background levels lead to substantial numbers of ions unevenly distributed in the drift volume. Even if a low ion backflow of the MPGD of one ion per primary electron is assumed the electrical field homogeneity degrades and hits on a track are deflected by more than 60 \textmu m from a straight drift path. For this reason TPCs traditionally feature a gating device, which can neutralize ions created in the gas amplification avalanche. The particular beam structure of the ILC consisting of 1312 bunches in a bunch train and a 199 ms break between bunch trains allows for a closed gate between bunch trains. To benefit fully from the MPGD approach of the readout modules, the standard multiwire approach for the gating device has been dismissed and a new approach of using a large aperture GEM-like structure is pursued \cite{GatingGEM}. Here, the electrical potential of the top and bottom electrode can be alternated between an 'open gate' and 'closed gate' voltage configuration. First samples of these devices have been produced and tested. They show both a good ion-stopping power and a high transparency for the primary electrons, which is important to retain the high performance. 
 
\section{Next Steps}
As the pad based approaches have demonstrated very similar results and final analysis of a first set of experiments are at a last stage, the groups plan a next generation with a near-final module design. Here, the gating device should be included in the design, a very similar pad layout should be used to ensure not only a good single track resolution, but also a good double-hit separation, a new readout electronics based on the sALTRO ASIC should be used and an integrated cooling concept, based either on 3D-printed cooling units or on Micro-channel pipes for the 2-phase CO$_2$.

For the GridPix design the next steps would also include a full module design with about 100 ASICs, a gating device and dedicated cooling foreseen. Here the challenges are more difficult, as the small device size requires a precise and very close positioning of the GridPixes on the module, to reduce the field distortions on the module and to maximize the fraction of active area, which is currently at 68.9 percent. Very promising new technology is the new Timepix4 ASIC which is almost a factor four larger and 4-side buttable with through silicon vias (TSV) which will allow to eliminate the wire bonding and to increase the density of the readout.

For all technologies also a substantial amount of simulation is needed to understand the impact of various parameter choices as for example pad sizes and electronics parameters. Besides, all of the technologies and the gating device have to be tested in a high magnetic field with $B=3.5$ T to confirm their performance under these conditions. In addition the reconstruction and analysis techniques have to be improved and sped up. In particular for the GridPix approach a partial online track reconstruction, possibly per ASIC and on the detector, are mandatory, as the amount of data becomes too large to handle otherwise.

\ack
This material is based upon work supported by the National Science Foundation of the U.S.A. under Grant No. 0935316 and was supported by the Japan Society for the Promotion of Science (JSPS) KAKENHI Grant No. 23000002. The research leading to these results has received funding from the European Commission under the 6th Framework Programme ``Structuring the European Research Area'', contract no. RII3-026126, and under the FP7 Research Infrastructures project AIDA, grant agreement no. 262025.
Special thanks go to Y. Makida, M. Kawai, K. Kasami and O. Araoka of the KEK IPNS cryogenic group, and A. Yamamoto of the KEK cryogenic centre for their support in the configuration and installation of the superconducting PCMAG solenoid.
The measurements leading to these results have been performed at the Test Beam Facility at DESY Hamburg (Germany), a member of the Helmholtz Association.
The collaboration would like to thank the technical team at the DESY II accelerator and test beam facility for the smooth operation of the test beam and the support during the test-beam campaign. The contributions to the experiment by the University of Lund, KEK, Nikhef and CEA are gratefully acknowledged.

\end{document}